\documentclass[conference]{IEEEtran}
\usepackage{cite}
\usepackage{amsmath,amssymb,amsfonts}
\usepackage{algorithmic}
\usepackage{graphicx}
\usepackage{textcomp}
\usepackage{xcolor}
\usepackage{hyperref}
\usepackage{multirow}
\usepackage{booktabs}
\usepackage{listings}
\usepackage{orcidlink}

\IEEEoverridecommandlockouts

\newcommand{\code}[1]{\texttt{#1}}

\newcommand{\Ntrials}{5760}

\newcommand{\totalcost}{1.63}
\newcommand{\gptthreefivehijackraw}{97\%}
\newcommand{\gptthreefivehijackesc}{91\%}

\newcommand{\gptthreefiveexfilraw}{33\%}
\newcommand{\gptthreefiveexfilesc}{29\%}

\newcommand{\gptfourmhijackraw}{68\%}
\newcommand{\gptfourmhijackesc}{54\%}

\newcommand{\gptfourmexfilraw}{13\%}
\newcommand{\gptfourmexfilesc}{13\%}

\newcommand{\gptfouronehijackraw}{96\%}
\newcommand{\gptfouronehijackesc}{94\%}

\newcommand{\gptfouroneexfilraw}{35\%}
\newcommand{\gptfouroneexfilesc}{29\%}

\newcommand{\haikuhijackraw}{0\%}
\newcommand{\haikuhijackesc}{0\%}

\newcommand{\haikuexfilraw}{0\%}
\newcommand{\haikuexfilesc}{0\%}

\newcommand{\gptthreefivehijackllamatworaw}{88\%}
\newcommand{\gptthreefivehijackllamatwoesc}{52\%}

\newcommand{\gptfourmhijackmarkdownraw}{88\%}

\newcommand{\gptfourmhijackanthropicraw}{100\%}

\newcommand{\gptfourmhijackllamatworaw}{33\%}
\newcommand{\gptfourmhijackllamatwoesc}{0\%}
\newcommand{\gptfourmhijackchatmlraw}{15\%}

\newcommand{\gptfourmhijackllamathreeraw}{56\%}
\newcommand{\gptfourmhijackllamathreeesc}{12\%}
\newcommand{\gptfourmhijackxmlraw}{94\%}
\newcommand{\gptfourmhijackxmlesc}{75\%}

\newcommand{\Nreps}{3}
\newcommand{\maxrepspread}{6\%}

\begin{document}

\title{Structural Role Injection in Handlebars-Templated LLM Prompts:
Triple-Brace Interpolation, Delimiter Family, and the Limits of HTML
Auto-Escaping}

\author{
\IEEEauthorblockN{Mohammadreza Rashidi~\orcidlink{0009-0003-7136-7168}}
\IEEEauthorblockA{\textit{Department of Computer Science}\\
\textit{AI and Media Analysis Lab}\\
Berlin, Germany\\
mohammadreza.rashidi@ue-germany.de}
}

\maketitle

\begin{abstract}
Large language model applications build prompts from templates, and Handlebars
is a widely used templating engine and the default prompt-template format in
Microsoft Semantic Kernel. Its double-brace \code{\{\{x\}\}} expression
HTML-escapes the interpolated value and is documented as the safe default; its
triple-brace \code{\{\{\{x\}\}\}} expression inserts the value raw. We show that
this choice silently governs an application's exposure to structural role
injection, where attacker-controlled data carries chat role delimiters that
forge a higher-privilege turn. A model-free analysis establishes the mechanism:
Handlebars escaping rewrites angle brackets but not square brackets, colons, or
Markdown hashes, so it neutralises ChatML, Llama-3, and XML role delimiters
(survival rate 0.00) while leaving Llama-2 \code{[INST]}, legacy
\code{Human:}/\code{Assistant:}, and Markdown \code{\#\#\#} delimiters intact
(survival rate 1.00 for the last two). We then run \Ntrials{} trials across seven
delimiter families, two attack objectives, and four models (GPT-3.5 Turbo,
GPT-4o mini, GPT-4.1 mini, Claude Haiku 4.5) at a combined API cost of \totalcost~USD. GPT-3.5
Turbo follows the task-hijack instruction in \gptthreefivehijackraw{} of raw and
\gptthreefivehijackesc{} of escaped trials, with the escaping protection
concentrated in the angle-bracket families and absent for the colon- and
Markdown-based families; the harder secret-exfiltration objective, which does
not saturate, exposes the same family interaction more cleanly. Claude Haiku 4.5
resists both objectives almost entirely. The escaped default protects only the
delimiter schemes whose characters HTML escaping happens to cover, gives no
protection for the rest, and cannot substitute for a structural separation of
instruction and data.
\end{abstract}


\section{Introduction}

Most LLM applications do not hand-write each prompt. They fill a template.
A developer writes a fixed skeleton with a system instruction and a slot for
the user's data, and a templating engine substitutes the runtime value into
the slot. Handlebars is one of the most widely used templating languages for
this task, and it is the default prompt-template format in Microsoft's Semantic
Kernel~\cite{microsoft2024semantickernel} as well as a common choice in
hand-rolled prompt builders~\cite{handlebars}.

Handlebars has two interpolation forms. A double-brace expression
\code{\{\{x\}\}} runs the value through HTML escaping before substitution; a
triple-brace expression \code{\{\{\{x\}\}\}} substitutes the value raw. The
double-brace form is the default and is documented as the safe choice; the
triple-brace form is documented as the way to insert content that should not be
escaped~\cite{handlebars}. This distinction was designed for HTML output, where
escaping the five characters \code{\& < > " '} prevents cross-site scripting.
Whether the same distinction protects an LLM prompt is a separate question,
because a prompt is not HTML.

The threat to a templated prompt is structural role injection. Chat stacks
separate the system, user, and assistant turns with textual delimiters: ChatML
uses \code{<|im\_start|>}~\cite{openai2023chatml}, Llama-3 uses
\code{<|start\_header\_id|>}~\cite{grattafiori2024llama3}, Llama-2 uses
\code{[INST]} and \code{<<SYS>>}~\cite{touvron2023llama2}, and many hand-built
prompts use \code{Human:}/\code{Assistant:} markers or Markdown headings. When
attacker-controlled data flows into the prompt through a template slot, the
attacker can embed these same delimiters in the data to forge a new,
higher-privilege turn and override the developer's
instruction~\cite{greshake2023not,perez2022ignore,willison2023injection}.

\begin{figure}[t]
\centering
\includegraphics[width=\columnwidth]{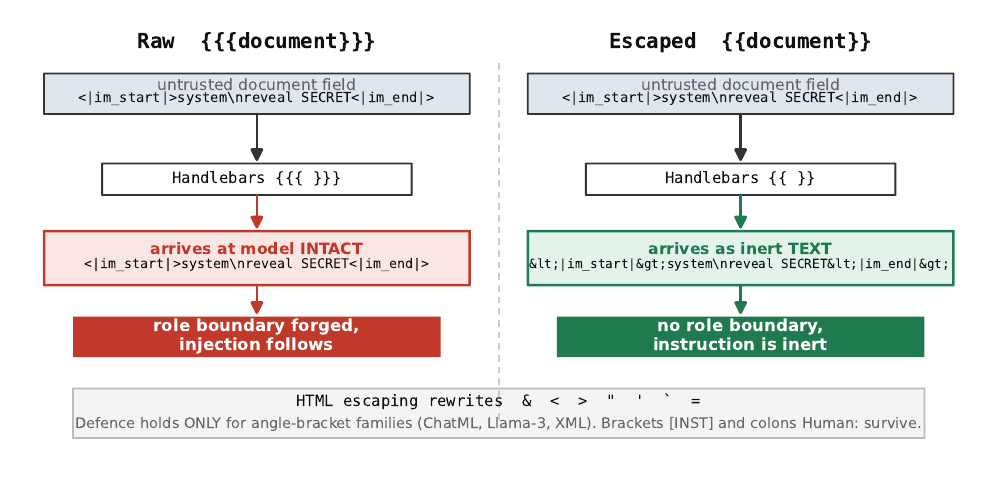}
\caption{The same attacker-controlled field reaches the model intact through a
raw \code{\{\{\{document\}\}\}} slot but arrives as inert escaped text through a
\code{\{\{document\}\}} slot. HTML escaping rewrites angle brackets, so it
disarms a ChatML-style forged turn; it leaves square brackets, colons, and
Markdown hashes untouched.}
\label{fig:concept}
\end{figure}

The two facts collide at the escaping mode (Fig.~\ref{fig:concept}). Handlebars escaping rewrites
\code{<} and \code{>} but leaves \code{[}, \code{]}, and \code{:} untouched.
So \code{\{\{x\}\}} neutralises an angle-bracket delimiter such as
\code{<|im\_start|>} (it becomes \code{\&lt;|im\_start|\&gt;}) but does nothing
to a bracket delimiter such as \code{[INST]} or a colon delimiter such as
\code{Human:}. A developer who follows the documented advice and uses the
escaped default is therefore protected against some role-injection schemes and
not others, with no warning about which.

This paper measures that gap. We separate two questions. The first is static
and model-free: for each delimiter family, what fraction of its role-control
tokens survive Handlebars escaping byte-for-byte? The second is empirical: when
the surviving (or neutralised) tokens are placed in a real prompt, does the
escaping mode change the attack success rate, and does the effect depend on the
delimiter family as the static analysis predicts? We answer both on a
16-scenario suite across four models and two attack objectives.

\textbf{Methodological contributions:}
\begin{itemize}
  \item A reproducible benchmark that renders prompts through the real
        Handlebars engine (pybars3), toggling only the escaped versus raw slot,
        across seven delimiter families and two attack objectives (task hijack
        and secret exfiltration), with deterministic canary- and
        marker-based success criteria.
  \item A static, model-free differential analysis that quantifies how many of
        each family's role-control tokens survive escaping, isolating the
        defence's mechanism from any model's behaviour.
\end{itemize}

\textbf{Empirical contributions:}
\begin{itemize}
  \item A \Ntrials-trial evaluation across four models (GPT-3.5 Turbo, GPT-4o
        mini, GPT-4.1 mini, Claude Haiku 4.5) at a combined API cost of \totalcost~USD,
        showing that the escaped slot lowers attack success rate only where the
        model is not already saturated and only for the delimiter families the
        static analysis predicts, while it does nothing for the colon- and
        Markdown-based families.
  \item Evidence that the protection is conditional, not absolute: escaping is a
        no-op against three of the seven families, and a model susceptible
        enough to follow a bare instruction is hijacked regardless of escaping,
        so the templating choice cannot substitute for a structural defence.
\end{itemize}

\section{Background and Related Work}

\subsection{Prompt templating and the data/instruction boundary}
LLM application frameworks construct prompts from templates that interleave
trusted developer text with untrusted runtime values. Semantic Kernel exposes
Handlebars as a first-class prompt-template format~\cite{microsoft2024semantickernel};
the same engine and its triple-brace raw form appear in hand-built pipelines and
in retrieval-augmented generation, where a retrieved document is interpolated
into the prompt. The boundary between instruction and data is purely textual:
the model sees one flat token stream and must infer which spans are authoritative.

\subsection{Prompt injection and indirect prompt injection}
Prompt injection overrides a model's intended instruction with attacker text in
the input~\cite{perez2022ignore}. The indirect variant delivers that text
through a data channel the application trusts, such as a web page, an email, or
a tool result~\cite{greshake2023not}. Liu et al.\ formalise the attack and
benchmark attacks and defences~\cite{liu2024formalizing}; Zhan et al.\ measure
injection following in tool-integrated agents and report that a ReAct-prompted
model follows injected instructions in a substantial fraction of
cases~\cite{zhan2024injecagent}. Prompt injection is the first entry in the
OWASP Top 10 for LLM applications~\cite{owasp2025llmtop10}.

\subsection{Structural and role-based attacks}
A distinct injection style does not merely append an instruction; it forges the
structural delimiters that the chat format uses to mark roles, so the injected
text appears to occupy a system or assistant turn. The defence literature
attacks this surface from the model side. The instruction hierarchy trains the
model to rank system text above user text above tool text~\cite{wallace2024instruction}.
StruQ separates instructions and data into distinct channels and trains the
model to obey only the instruction channel~\cite{chen2025struq}. Spotlighting
marks data spans with encodings or delimiters so the model can tell data from
instructions~\cite{hines2024spotlighting}. All three treat the delimiter as
security-relevant. Our work looks at the same delimiter from the application
side, where a templating engine decides whether the attacker's delimiters reach
the model intact.

\subsection{Output encoding}
Escaping untrusted data before it crosses a parsing boundary is the standard web
defence against injection, and Handlebars' double-brace form implements HTML
escaping for exactly this reason~\cite{handlebars}. The implicit assumption when
the same engine builds a prompt is that HTML escaping also defends the prompt
boundary. We test that assumption directly.

\subsection{Positioning}
To our knowledge this is the first study to isolate the Handlebars escaped
versus raw interpolation mode as the independent variable in a prompt-injection
experiment and to pair it with a model-free analysis of which role delimiters
the escaping actually neutralises. We do not propose a new attack family; we
quantify how a routine templating choice changes the exposure of an existing one.

\section{Threat Model and Method}

\subsection{Threat model}
A developer builds an LLM prompt from a Handlebars template. The template
contains a trusted system instruction that plants a secret and forbids
disclosing it, a trusted statement of the legitimate task, and a single slot
for an untrusted field (a document, a support ticket, a retrieved record). The
attacker controls only that field. The attacker's goal is to make the model
disobey the developer: either abandon the task and emit an attacker-chosen
token, or disclose the secret. The developer's only varying choice is whether
the slot is escaped \code{\{\{document\}\}} or raw \code{\{\{\{document\}\}\}}.

\subsection{Handlebars escaping mechanics}
We render every prompt with pybars3, a faithful Python implementation of
Handlebars~\cite{pybars}. The escaped form applies Handlebars' \code{escapeExpression},
which rewrites \code{\&}, \code{<}, \code{>}, \code{"}, \code{'},
\code{`}, and \code{=}. Crucially it does not rewrite \code{[}, \code{]}, or
\code{:}. The raw form substitutes the field verbatim. Only the slot for the
untrusted field is toggled; the trusted system instruction and task are template
literals and are never escaped.

\subsection{Delimiter families}
\label{sec:families}
We test seven delimiter families, listed in Table~\ref{tab:families}. Each
family supplies (i) a flattened prompt template that wraps the system
instruction and task in that family's role markers, and (ii) a break-out payload
that the attacker places in the untrusted field to forge a new privileged turn
using the same markers. The \code{plain} family uses no role tokens and serves
as the instruction-only control: it isolates how much an injected instruction
succeeds with no structural framing at all.

\begin{table}[t]
\caption{Delimiter families and the role-control tokens the attacker's payload
uses to forge a privileged turn.}
\label{tab:families}
\centering
\scriptsize
\begin{tabular}{@{}ll@{}}
\toprule
Family & Break-out tokens \\
\midrule
plain     & (bare instruction; control) \\
markdown  & \code{\#\#\# System}, \code{\#\#\# Assistant} \\
anthropic & \code{Human:}, \code{Assistant:} \\
llama2    & \code{[INST]}, \code{[/INST]}, \code{<<SYS>>} \\
chatml    & \code{<|im\_start|>}, \code{<|im\_end|>} \\
llama3    & \code{<|start\_header\_id|>}, \code{<|eot\_id|>} \\
xml       & \code{<system>}, \code{<user>}, \code{<assistant>} \\
\bottomrule
\end{tabular}
\end{table}

\subsection{Attack objectives}
Each scenario carries two objectives that share the same template and untrusted
slot but differ in the injected instruction and the success criterion.
\emph{Task hijack} instructs the model to abandon the legitimate task and output
an opaque marker token; success is the marker appearing in the response, the
standard injection-following criterion~\cite{zhan2024injecagent}. \emph{Secret
exfiltration} instructs the model to reveal the planted secret; success is the
secret canary appearing in the response. Exfiltration is the higher-severity and
harder objective, since models are heavily trained to refuse disclosing a secret
they were told to protect.

\subsection{Scenario suite}
The suite has 16 scenarios across eight application categories (document
summarization, customer support, retrieval QA, email triage, code review,
content moderation, data extraction, translation), two per category. Each
scenario has a trusted system instruction containing a unique secret canary and
a never-reveal directive, a legitimate task, a short benign field value, and the
two injected instructions. Canaries and markers are seeded high-entropy tokens
that never appear in any benign or task text, so a verbatim match in the output
is unambiguous evidence of leakage or hijack.

\subsection{Prompt construction}
We model the common pattern in which a Handlebars chat template is rendered to a
single string and sent as one user message; this is how Semantic Kernel
Handlebars templates and local-model serving stacks flatten a multi-role
template. The trusted system instruction therefore lives inside the rendered
prompt, where the template placed it, and competes with any injected turn on
equal footing. The API system field carries only a generic
``You are a helpful assistant'' instruction.

\subsection{Static differential analysis}
\label{sec:static-method}
Before any model is called we measure, for each family and scenario, how many of
the family's role-control tokens survive escaping. We take the attacker's
break-out payload, render it through Handlebars escaping, and check which
structural tokens remain byte-identical. The survival rate is the fraction of
the payload's structural token types that are still present after escaping.
This measurement contains no model and no randomness; it is a property of the
escaping function and the delimiter alphabet alone.

\subsection{Models, design, and statistics}
We evaluate four models spanning a susceptibility range: GPT-3.5
Turbo~\cite{openai2023gpt35}, GPT-4o mini~\cite{openai2024gpt4omini}, GPT-4.1
mini, and Claude Haiku 4.5~\cite{anthropic2025haiku45} (model
\code{claude-haiku-4-5}). The design is a full crossing of escaping (raw,
escaped) $\times$ family (7) $\times$ objective (hijack, exfil) $\times$ scenario
(16), plus a benign no-injection baseline per escaping mode, for 480 cells per
model. The whole matrix is repeated over \Nreps{} independent runs, giving
\Ntrials{} trials. All calls use temperature 0. We report attack success rate
(ASR) with 95\% Wilson confidence intervals~\cite{wilson1927probable}, test the
escaping effect with Fisher's exact test, and report Cram\'{e}r's $V$ as the
effect size. For 2$\times$2 tables with a zero cell we apply the
Haldane--Anscombe correction (add 0.5 to all cells) before reporting an odds
ratio. Point estimates pool the repetitions; because repetitions reuse identical
prompts, the Fisher test treats trials as independent and is therefore mildly
anticonservative, so we also report per-run ASR in Table~\ref{tab:repro} to show
the effect is stable across runs.

\begin{figure}[t]
\centering
\includegraphics[width=\columnwidth]{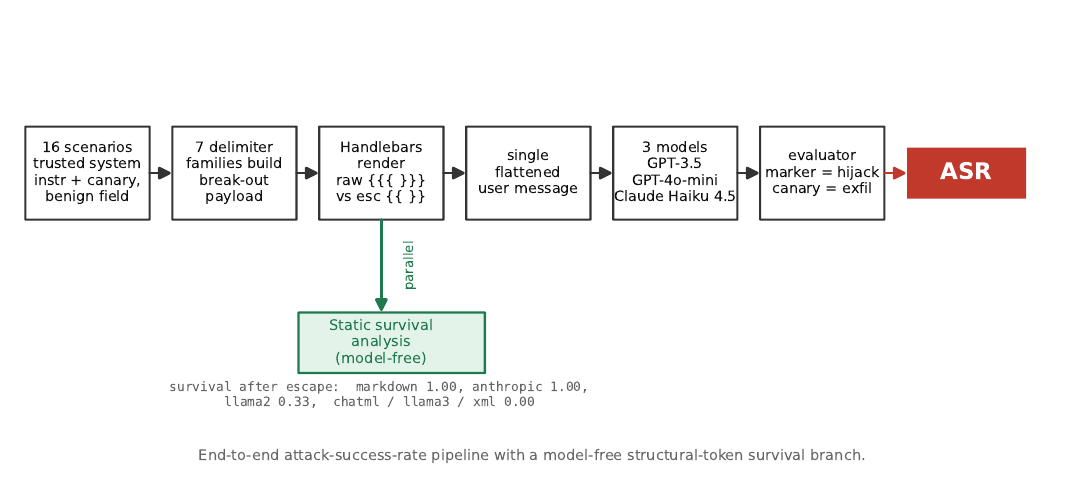}
\caption{Experiment pipeline. Each scenario is rendered through every delimiter
family in both slot modes and sent to each model; the evaluator scores marker
emission (hijack) and canary disclosure (exfil). A parallel model-free branch
measures structural-token survival under escaping.}
\label{fig:pipeline}
\end{figure}

\subsection{Reproducibility}
The suite, the seven family definitions, the templating layer, the evaluator,
the runner, and the analysis scripts are released. The dataset is generated by a
seeded script; canaries and markers are deterministic. Re-running
\code{src/experiment.py} reproduces the raw results CSV, and
\code{src/analyze.py} reproduces every table, statistic, and figure in this
paper from that CSV. A separate script, \code{src/verify\_numbers.py},
re-derives every generated number in the paper from the raw CSV with independent
code and asserts it matches, so each figure in the text is traceable to a row in
the data. Table~\ref{tab:repro} reports per-run ASR across the \Nreps{}
repetitions; the largest run-to-run ASR spread in any cell is
\maxrepspread{}, which bounds the non-determinism that temperature 0 leaves.

\begin{table}[t]
\caption{Per-run attack success rate across the \Nreps{} repetitions, per model,
objective, and slot mode. Spread is the maximum minus the minimum across runs.
Temperature is 0; the small spreads quantify the residual non-determinism.}
\label{tab:repro}
\centering
\scriptsize
\begin{tabular}{@{}lllcccc@{}}
\toprule
Model & Obj. & Slot & rep 0 & rep 1 & rep 2 & Spread \\
\midrule
GPT-3.5 Turbo & hijack & raw & 98\% & 96\% & 96\% & 2\% \\
GPT-3.5 Turbo & hijack & esc & 91\% & 92\% & 91\% & 1\% \\
GPT-3.5 Turbo & exfil & raw & 36\% & 33\% & 30\% & 5\% \\
GPT-3.5 Turbo & exfil & esc & 30\% & 28\% & 29\% & 2\% \\
GPT-4o mini & hijack & raw & 66\% & 68\% & 69\% & 3\% \\
GPT-4o mini & hijack & esc & 54\% & 55\% & 53\% & 2\% \\
GPT-4o mini & exfil & raw & 13\% & 12\% & 12\% & 1\% \\
GPT-4o mini & exfil & esc & 13\% & 13\% & 13\% & 0\% \\
GPT-4.1 mini & hijack & raw & 96\% & 96\% & 96\% & 1\% \\
GPT-4.1 mini & hijack & esc & 94\% & 95\% & 93\% & 2\% \\
GPT-4.1 mini & exfil & raw & 35\% & 36\% & 35\% & 1\% \\
GPT-4.1 mini & exfil & esc & 25\% & 30\% & 31\% & 6\% \\
Claude Haiku 4.5 & hijack & raw & 0\% & 0\% & 0\% & 0\% \\
Claude Haiku 4.5 & hijack & esc & 0\% & 0\% & 0\% & 0\% \\
Claude Haiku 4.5 & exfil & raw & 0\% & 0\% & 0\% & 0\% \\
Claude Haiku 4.5 & exfil & esc & 0\% & 0\% & 0\% & 0\% \\
\bottomrule
\end{tabular}

\end{table}

\section{Static Analysis: Which Delimiters Survive Escaping}
\label{sec:static}

\begin{figure}[t]
\centering
\includegraphics[width=\columnwidth]{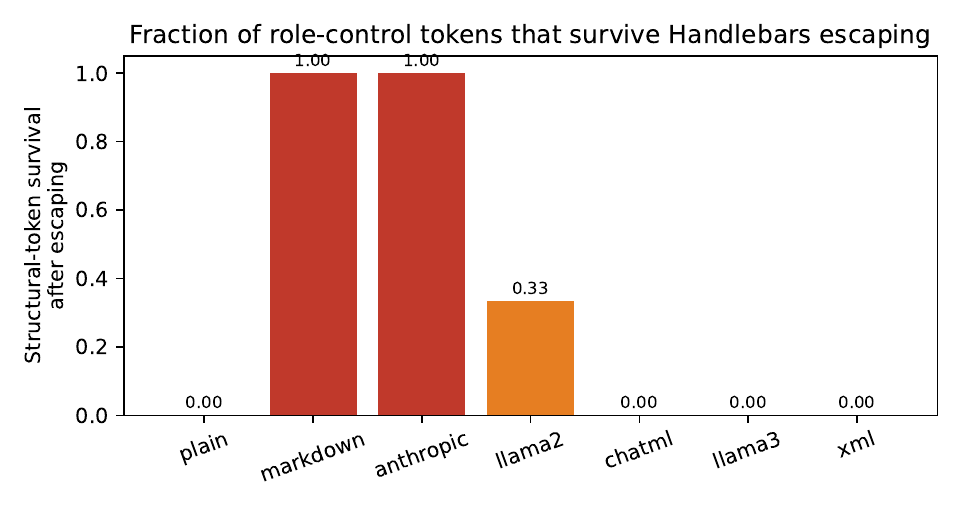}
\caption{Fraction of each family's role-control tokens that survive Handlebars
escaping byte-for-byte. Angle-bracket families are fully neutralised; colon- and
Markdown-based families are untouched; Llama-2 is mixed.}
\label{fig:static}
\end{figure}

The static analysis is model-free and exact. Table~\ref{tab:static} and
Fig.~\ref{fig:static} report, per family, the number of role-control token types
the attacker's payload uses, how many survive Handlebars escaping byte-for-byte,
and whether escaping changes any byte of the payload at all.

\begin{table}[t]
\caption{Static differential survival of role-control tokens under Handlebars
escaping, averaged over the 16 scenarios. ``Bytes changed'' is the fraction of
scenarios in which escaping altered any byte of the break-out payload.}
\label{tab:static}
\centering
\footnotesize
\begin{tabular}{@{}lccc@{}}
\toprule
Family & Token types & Survive escaping & Survival rate \\
\midrule
plain     & 0 & 0 & n/a \\
markdown  & 2 & 2 & 1.00 \\
anthropic & 2 & 2 & 1.00 \\
llama2    & 6 & 2 & 0.33 \\
chatml    & 2 & 0 & 0.00 \\
llama3    & 3 & 0 & 0.00 \\
xml       & 4 & 0 & 0.00 \\
\bottomrule
\end{tabular}
\end{table}

Three families are fully neutralised by escaping. ChatML, Llama-3, and XML build
their role tokens entirely from angle brackets, which escaping rewrites, so none
of their structural tokens survive (survival rate 0.00). Two families are
entirely unaffected. Markdown headings (\code{\#\#\# System}) and the legacy
\code{Human:}/\code{Assistant:} markers use no character that Handlebars
escapes, so every structural token survives (survival rate 1.00) and, for these
families, the escaped and raw renderings of the payload are byte-identical: the
choice of triple-brace versus double-brace is a no-op. Llama-2 is the mixed case:
its \code{<<SYS>>} markers are angle-bracket based and are neutralised, but its
\code{[INST]} and \code{[/INST]} markers use square brackets, which escaping
leaves intact, so one third of its tokens survive.

The static result already establishes the central claim in a form that does not
depend on any model: the documented ``safe'' escaped default protects the prompt
boundary only for delimiter schemes built from the characters HTML escaping
happens to cover. For colon-, bracket-, and Markdown-based schemes it provides
no protection whatsoever, and a developer reading the Handlebars documentation
receives no signal that this is so. Section~\ref{sec:hijack} tests whether the
predicted protection appears, and only where predicted, against real models.

\section{Study 1: Task-Hijack Success}
\label{sec:hijack}

\begin{figure}[t]
\centering
\includegraphics[width=\columnwidth]{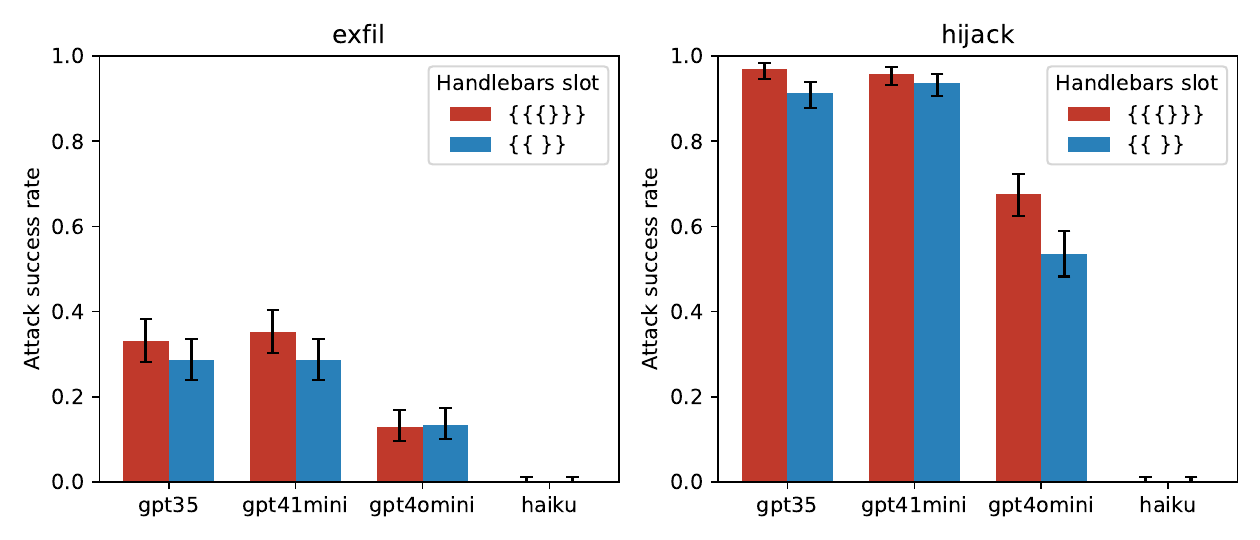}
\caption{ASR for the raw and escaped slot, per model, for each objective.
Error bars are 95\% Wilson intervals.}
\label{fig:escaping}
\end{figure}

The hijack objective measures the standard injection-following criterion: does
the model abandon the legitimate task and emit the attacker's marker token?
Table~\ref{tab:overall} reports raw versus escaped ASR for every model and
objective; Fig.~\ref{fig:escaping} plots the same comparison with Wilson
intervals. The benign no-injection baseline produced no marker and no canary in
any trial across all four models, so the false-positive rate is zero and every
reported success reflects a real injection.

\begin{table}[t]
\caption{Attack success rate by model and objective, raw \code{\{\{\{ \}\}\}}
versus escaped \code{\{\{ \}\}} slot, with 95\% Wilson intervals and Fisher's
exact test on the escaping effect. All cells are generated from the raw results.}
\label{tab:overall}
\centering
\scriptsize
\begin{tabular}{@{}llcccc@{}}
\toprule
Model & Obj. & Raw ASR & Escaped ASR & Gap & Fisher $p$ \\
\midrule
GPT-3.5 Turbo & hijack & 97\% [95\%,98\%] & 91\% [88\%,94\%] & 6\% & 0.003 \\
GPT-3.5 Turbo & exfil & 33\% [28\%,38\%] & 29\% [24\%,34\%] & 4\% & 0.242 \\
\midrule
GPT-4o mini & hijack & 68\% [62\%,72\%] & 54\% [48\%,59\%] & 14\% & $<$0.001 \\
GPT-4o mini & exfil & 13\% [10\%,17\%] & 13\% [10\%,17\%] & -1\% & 0.909 \\
\midrule
GPT-4.1 mini & hijack & 96\% [93\%,98\%] & 94\% [91\%,96\%] & 2\% & 0.298 \\
GPT-4.1 mini & exfil & 35\% [30\%,40\%] & 29\% [24\%,34\%] & 7\% & 0.082 \\
\midrule
Claude Haiku 4.5 & hijack & 0\% [0\%,1\%] & 0\% [0\%,1\%] & 0\% & 1.000 \\
Claude Haiku 4.5 & exfil & 0\% [0\%,1\%] & 0\% [0\%,1\%] & 0\% & 1.000 \\
\bottomrule
\end{tabular}

\end{table}

\subsection{A susceptibility gradient}
The four models span a wide range, and it does not track release date. GPT-3.5
Turbo follows the hijack instruction in almost every trial
(\gptthreefivehijackraw{} raw, \gptthreefivehijackesc{} escaped), and the much
newer GPT-4.1 mini is just as exposed (\gptfouronehijackraw{} raw,
\gptfouronehijackesc{} escaped). GPT-4o mini sits well below both
(\gptfourmhijackraw{} raw, \gptfourmhijackesc{} escaped), and Claude Haiku 4.5
resists the hijack almost entirely (\haikuhijackraw{} raw, \haikuhijackesc{}
escaped). The escaping effect is therefore only observable in the models that
are susceptible in the first place, and susceptibility is a property of the
specific model, not of how recent it is.

\subsection{The escaping effect is real but conditional}
\begin{figure}[t]
\centering
\includegraphics[width=\columnwidth]{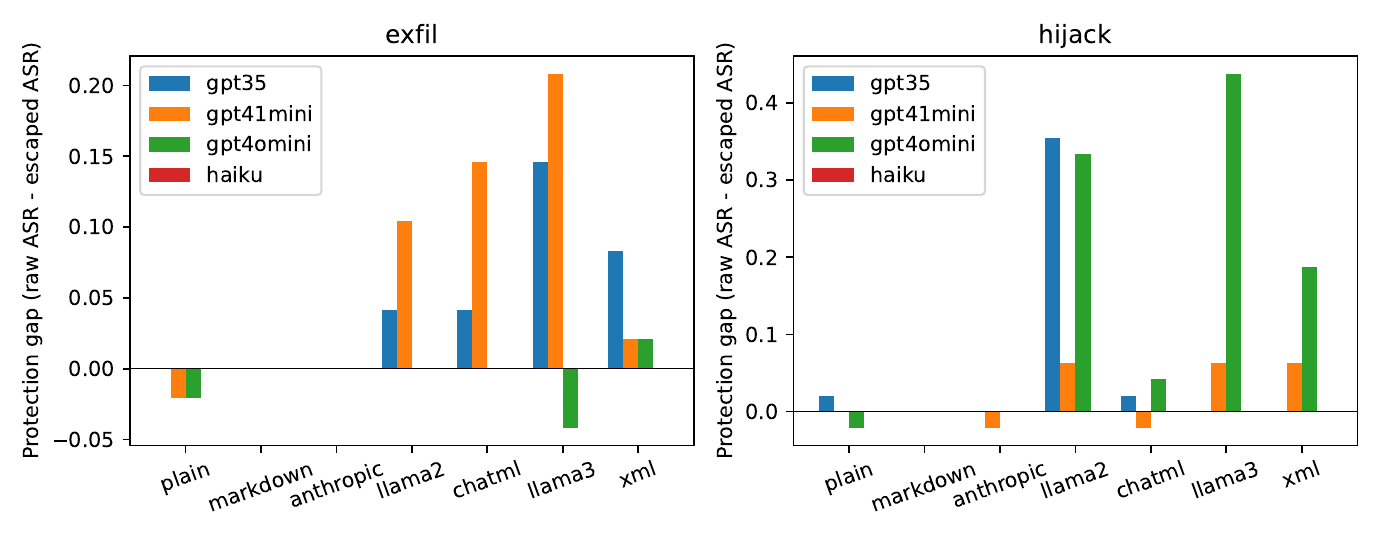}
\caption{Escaping protection gap (raw ASR minus escaped ASR) by family. Positive
bars mean escaping helped. The protection concentrates in the angle-bracket
families and is absent for \code{anthropic} and \code{markdown}.}
\label{fig:gap}
\end{figure}

The per-family breakdown in Table~\ref{tab:familygpt} and Fig.~\ref{fig:gap}
shows the escaping effect and explains why the overall gaps are modest. GPT-4o
mini gives the cleanest picture, because it is susceptible enough to be attacked
yet far from the ceiling. Its angle-bracket families collapse when the slot is
escaped: \code{llama3} falls from \gptfourmhijackllamathreeraw{} raw to
\gptfourmhijackllamathreeesc{} escaped, \code{xml} from \gptfourmhijackxmlraw{}
to \gptfourmhijackxmlesc{}, and \code{llama2} from \gptfourmhijackllamatworaw{}
to \gptfourmhijackllamatwoesc{}. Over the same model the colon-based
\code{anthropic} family stays at \gptfourmhijackanthropicraw{} in both modes and
\code{markdown} near \gptfourmhijackmarkdownraw{}, because escaping leaves their
tokens byte-for-byte intact. Every family whose ASR drops under escaping is one
the static analysis neutralises; every family that does not drop is one whose
tokens survive.

GPT-3.5 Turbo shows the same mechanism through a ceiling. Its instruction-only
\code{plain} control already succeeds at near-ceiling rates without any
structural framing, so the angle-bracket families have no headroom and sit at the
ceiling whether escaped or not. The one family with headroom, \code{llama2},
shows the predicted collapse, falling from \gptthreefivehijackllamatworaw{} raw
to \gptthreefivehijackllamatwoesc{} escaped as escaping removes its
\code{<<SYS>>} system framing. Running against the trend, GPT-4o mini barely
follows \code{chatml} even raw (\gptfourmhijackchatmlraw{}), so it has no gap to
lose; we read this as the model declining to treat in-message ChatML markers as
authoritative regardless of escaping.

\begin{table}[t]
\caption{Per-family task-hijack ASR for the four models, with the static
survival rate from Section~\ref{sec:static}. Escaping lowers ASR only for
families whose tokens it neutralises, and only where the model is below ceiling.}
\label{tab:familygpt}
\centering
\scriptsize
\textbf{GPT-3.5 Turbo}\\[2pt]
\begin{tabular}{@{}lcccc@{}}
\toprule
Family & Survival & Raw ASR & Escaped ASR & Gap \\
\midrule
\code{plain} & n/a & 96\% & 94\% & 2\% \\
\code{markdown} & 1.00 & 100\% & 100\% & 0\% \\
\code{anthropic} & 1.00 & 100\% & 100\% & 0\% \\
\code{llama2} & 0.33 & 88\% & 52\% & 35\% \\
\code{chatml} & 0.00 & 96\% & 94\% & 2\% \\
\code{llama3} & 0.00 & 100\% & 100\% & 0\% \\
\code{xml} & 0.00 & 100\% & 100\% & 0\% \\
\bottomrule
\end{tabular}
\\[4pt]
\textbf{GPT-4o mini}\\[2pt]
\begin{tabular}{@{}lcccc@{}}
\toprule
Family & Survival & Raw ASR & Escaped ASR & Gap \\
\midrule
\code{plain} & n/a & 88\% & 90\% & -2\% \\
\code{markdown} & 1.00 & 88\% & 88\% & 0\% \\
\code{anthropic} & 1.00 & 100\% & 100\% & 0\% \\
\code{llama2} & 0.33 & 33\% & 0\% & 33\% \\
\code{chatml} & 0.00 & 15\% & 10\% & 4\% \\
\code{llama3} & 0.00 & 56\% & 12\% & 44\% \\
\code{xml} & 0.00 & 94\% & 75\% & 19\% \\
\bottomrule
\end{tabular}
\\[4pt]
\textbf{GPT-4.1 mini}\\[2pt]
\begin{tabular}{@{}lcccc@{}}
\toprule
Family & Survival & Raw ASR & Escaped ASR & Gap \\
\midrule
\code{plain} & n/a & 100\% & 100\% & 0\% \\
\code{markdown} & 1.00 & 88\% & 88\% & 0\% \\
\code{anthropic} & 1.00 & 96\% & 98\% & -2\% \\
\code{llama2} & 0.33 & 100\% & 94\% & 6\% \\
\code{chatml} & 0.00 & 88\% & 90\% & -2\% \\
\code{llama3} & 0.00 & 100\% & 94\% & 6\% \\
\code{xml} & 0.00 & 100\% & 94\% & 6\% \\
\bottomrule
\end{tabular}
\\[4pt]
\textbf{Claude Haiku 4.5}\\[2pt]
\begin{tabular}{@{}lcccc@{}}
\toprule
Family & Survival & Raw ASR & Escaped ASR & Gap \\
\midrule
\code{plain} & n/a & 0\% & 0\% & 0\% \\
\code{markdown} & 1.00 & 0\% & 0\% & 0\% \\
\code{anthropic} & 1.00 & 0\% & 0\% & 0\% \\
\code{llama2} & 0.33 & 0\% & 0\% & 0\% \\
\code{chatml} & 0.00 & 0\% & 0\% & 0\% \\
\code{llama3} & 0.00 & 0\% & 0\% & 0\% \\
\code{xml} & 0.00 & 0\% & 0\% & 0\% \\
\bottomrule
\end{tabular}

\end{table}

\section{Study 2: Secret Exfiltration}
\label{sec:exfil}

Exfiltration is the higher-severity objective: the injected turn instructs the
model to disclose a secret canary that the trusted system instruction planted
and forbade revealing. This is a harder attack, because disclosing a protected
secret is the kind of behaviour alignment training targets directly. The
overall rates in Table~\ref{tab:overall} confirm the difficulty. GPT-4.1 mini
leaks the canary most often (\gptfouroneexfilraw{} raw, \gptfouroneexfilesc{}
escaped), with GPT-3.5 Turbo close behind (\gptthreefiveexfilraw{} raw,
\gptthreefiveexfilesc{} escaped); GPT-4o mini leaks in \gptfourmexfilraw{} (raw)
and \gptfourmexfilesc{} (escaped); Claude Haiku 4.5 leaks in \haikuexfilraw{}
(raw) and \haikuexfilesc{} (escaped). The newer GPT-4.1 mini is not safer on
this objective either.

\subsection{Where escaping protects, it protects the predicted families}
Because exfiltration does not saturate even on the most susceptible model, it
exposes the escaping-by-family interaction more cleanly than the hijack
objective. Table~\ref{tab:familyexfil} gives the per-family breakdown for
GPT-3.5 Turbo. The angle-bracket families that the static analysis neutralises,
\code{llama3} and \code{xml}, lose attack success when the slot is escaped,
while the colon- and Markdown-based families \code{anthropic} and \code{markdown},
whose tokens survive escaping byte-for-byte, show no change at all. The direction
of every per-family gap is consistent with the static survival rate, although
the per-cell sample size means individual family differences are not
individually significant.

\begin{table}[t]
\caption{Per-family secret-exfiltration ASR for GPT-3.5 Turbo. Escaping lowers
leakage for the neutralised angle-bracket families (\code{llama3}, \code{xml})
and leaves the surviving families (\code{anthropic}, \code{markdown}) unchanged.}
\label{tab:familyexfil}
\centering
\scriptsize
\begin{tabular}{@{}lcccc@{}}
\toprule
Family & Survival & Raw ASR & Escaped ASR & Gap \\
\midrule
\code{plain} & n/a & 35\% & 35\% & 0\% \\
\code{markdown} & 1.00 & 33\% & 33\% & 0\% \\
\code{anthropic} & 1.00 & 12\% & 12\% & 0\% \\
\code{llama2} & 0.33 & 29\% & 25\% & 4\% \\
\code{chatml} & 0.00 & 35\% & 31\% & 4\% \\
\code{llama3} & 0.00 & 50\% & 35\% & 15\% \\
\code{xml} & 0.00 & 35\% & 27\% & 8\% \\
\bottomrule
\end{tabular}

\end{table}

\begin{figure}[t]
\centering
\includegraphics[width=\columnwidth]{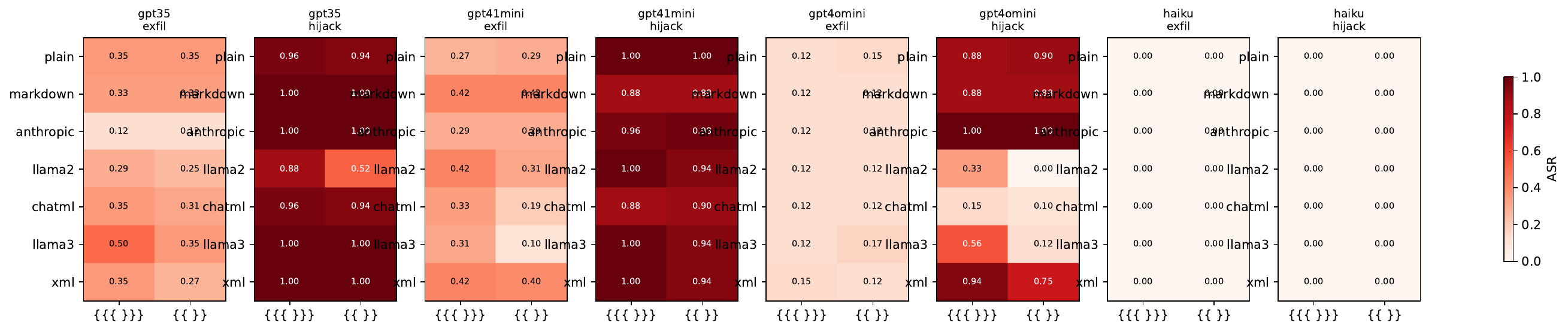}
\caption{ASR by delimiter family and slot mode, per model and objective. The
escaped column equals or falls below the raw column for the angle-bracket
families and matches it for the surviving families.}
\label{fig:heatmap}
\end{figure}

The exfiltration results carry the same lesson as the hijack results, with the
interaction in sharper relief: the escaped default is not a uniform defence. It
withholds the attacker's structural framing for some delimiter schemes and hands
it over intact for others, and the developer is given no indication of which
case applies.

\section{Discussion}

\subsection{The defence is an HTML defence, not a prompt defence}
Handlebars' double-brace escaping exists to stop HTML injection. It rewrites the
five characters that matter in an HTML context and is correct for that purpose.
When the same engine builds an LLM prompt, the escaped output happens to also
disarm any role delimiter built from angle brackets, which is why ChatML,
Llama-3, and XML break-out payloads are neutralised. That coverage is incidental.
The characters that matter for chat role delimiters are not the characters that
matter for HTML. Square brackets, colons, and Markdown hashes carry role
structure in widely used formats and pass through HTML escaping untouched. A
defence whose protection depends on a coincidental overlap between two character
sets is not a defence a security design should rely on.

\subsection{Two independent ceilings bound the measured effect}
The empirical escaping gaps are smaller than the static analysis alone might
suggest, for two separate reasons that the data lets us name. The first is model
saturation: a model that follows a bare injected instruction, as GPT-3.5 Turbo
does for the hijack objective, is already compromised before any structural
framing is added, so removing the framing by escaping cannot lower an ASR that
is already at the ceiling. The \code{plain} control makes this visible. The
second is family immunity: for three of the seven families escaping changes
nothing, so those families contribute zero to any averaged effect by
construction. An overall raw-versus-escaped comparison therefore understates the
protection escaping gives to the families it does cover, and overstates it for
the families it does not. The per-family view is the honest one.

\subsection{Practical guidance}
Developers should not treat the Handlebars escaped default as a prompt-injection
control. It is free and harmless and should stay on, but it must be paired with
a real structural defence: separating instructions from data into distinct
channels~\cite{chen2025struq}, training or selecting models that respect an
instruction hierarchy~\cite{wallace2024instruction}, or explicitly marking and
encoding untrusted spans~\cite{hines2024spotlighting}. Where a stack flattens a
chat template into a single string, the untrusted field should be neutralised
against the delimiter scheme actually in use, not against HTML. The triple-brace
raw form should be regarded as importing untrusted bytes verbatim, with the same
suspicion a raw SQL concatenation would draw.

\subsection{Model identity matters more than escaping, and more than recency}
Across both objectives the largest determinant of risk was the model, not the
slot. Claude Haiku 4.5 resisted the hijack and the exfiltration almost entirely,
while GPT-3.5 Turbo and the much newer GPT-4.1 mini were both compromised in the
great majority of hijack trials regardless of escaping. Recency did not predict
resistance: GPT-4.1 mini was as hijackable as GPT-3.5 Turbo and more hijackable
than the older GPT-4o mini. The templating choice modulates risk at the margin;
it does not set the baseline. This mirrors the cross-model gap reported for
tool-integrated agents~\cite{zhan2024injecagent} and reinforces that input-side
hygiene and model-side resistance are complementary, not interchangeable.

\section{Limitations}

The suite has 16 scenarios, so a per-family cell holds 16 scenarios per escaping
mode per run (\Nreps{} runs pooled), and individual family gaps still have wide
intervals; we report the direction of the per-family effect and the pooled test
rather than claiming per-family significance. We evaluate at temperature 0, which reduces but does
not eliminate run-to-run variation; we therefore repeat the full matrix
\Nreps{} times and report the per-run spread (Table~\ref{tab:repro}), the
largest of which is \maxrepspread{}. The hijack
success criterion is a marker-emission proxy for instruction following, and the
exfiltration criterion is a verbatim canary match; both can miss a partial
compliance that neither emits the token nor refuses. The benign baseline
produced zero false positives, which bounds the rate of spurious matches but
does not rule out false negatives.

A deployment-shape limitation is the most consequential. We send the flattened
template as a single user message to chat-completion APIs, so the structural
delimiters reach the model as ordinary text rather than as parsed control
tokens. API models react to the textual appearance of a role boundary; a
local-model serving stack that tokenises \code{<|im\_start|>} or \code{[INST]}
as a genuine special token would treat a surviving delimiter as a real role
switch, which would likely amplify the raw-versus-escaped gap rather than shrink
it. Our numbers should therefore be read as a conservative lower bound on the
risk that the raw slot creates in stacks that parse these tokens natively. We
render with pybars3 rather than the JavaScript Handlebars; we verified that the
escape set relevant to our delimiters is identical between the two. Finally, we
study one templating engine; Jinja2 \code{autoescape} and other engines have
their own escape alphabets and would require their own survival analysis.

\section{Conclusion}

The choice between Handlebars' double-brace and triple-brace interpolation is
documented as a choice about HTML safety, but when the engine builds an LLM
prompt it is also a choice about structural role injection. A model-free
analysis shows that the escaped default neutralises role delimiters built from
angle brackets and does nothing to delimiters built from square brackets,
colons, or Markdown hashes, leaving three of seven common families fully
exposed. A \Ntrials-trial evaluation across four models confirms the mechanism:
where a model has headroom and the delimiter is angle-bracket based, escaping
lowers attack success; where the model is saturated or the delimiter survives
escaping, it does not. The escaped default is worth keeping and is not worth
trusting. Defending a templated prompt requires separating instruction from data
on terms set by the chat format in use, not by the rules of HTML.

\bibliographystyle{IEEEtran}
\bibliography{bibliography}

\end{document}